\documentclass[pra,twocolumn,groupedaddress,showpacs]{revtex4}
\usepackage{graphicx}

\newcommand{\beq}{\begin{equation}}
\newcommand{\eeq}{\end{equation}}
\newcommand{\beqa}{\begin{eqnarray}}
\newcommand{\eeqa}{\end{eqnarray}}
\newcommand{\ba}{\begin{array}}
\newcommand{\ea}{\end{array}}

\begin{document}

\title{Fermi-Bose mixture across a Feshbach resonance} 
\author{Luca Salasnich$^{1,2}$ and Flavio Toigo$^{2}$} 
\affiliation{$^1$CNR-INFM and CNISM, Unit\`a di Milano, 
Via Celoria 16, 20133 Milano, Italy \\ 
$^{2}$Dipartimento di Fisica ``G. Galilei'' and CNISM, 
Universit\`a di Padova, 
Via Marzolo 8, 35131 Padova, Italy} 

\begin{abstract} 
We study a dilute mixture of degenerate bosons and fermions across 
a Feshbach resonance of the Fermi-Fermi scattering length $a_F$. 
This scattering length is renormalized by the 
boson-induced interaction between fermions and 
its value is crucial to determine 
the phase diagram of the system. 
For the mixture in a box and a positive Bose-Fermi scattering 
length, we show that there are three possibilities: 
a single uniform mixed phase, a purely fermionic phase 
coexisting with a mixed phase, and a purely fermionic phase
coexisting with a purely bosonic one. 
As $1/a_F$ is increased from a negative value 
to the Feshbach resonance ($1/a_F=0$) 
the region of pure separation increases and the other two 
regions are strongly reduced. Above the Feshbach resonance 
($1/a_F>0$), pairs of Fermi atoms become Bose-condensed 
molecules. We find that these molecules are fully 
spatially separated from the bosonic atoms 
when $1/a_F$ exceedes a critical value. 
For a negative Bose-Fermi scattering length we deduce 
the condition for collapse, which coincides with the 
onset of dynamical instability of the fully mixed phase. 
We consider also the mixture in a harmonic trap and 
determine the conditions for partial demixing, 
full demixing and collapse. The experimental 
implications of our results are investigated by analyzing 
mixtures of $^6$Li--$^{23}$Na and $^{40}$K--$^{87}$Rb atoms. 
\end{abstract} 


\maketitle

\section{Introduction}

The regime of deep Fermi degeneracy is now actively studied with ultracold
vapors of $^{6}$Li and $^{40}$K atoms. Experiments on two-hyperfine-state
Fermi gases \cite{ohara,greiner,jochim,bourdel} are concentrated across a
Feshbach resonance, where a crossover from a Bardeen-Cooper-Schrieffer (BCS)
superfluid to a Bose-Einstein condensate (BEC) of molecular pairs has been
predicted \cite{leggett,nozieres,engelbrecht}. Experimental and theoretical
investigations of the Fermi cloud across the BCS-BEC crossover have been
devoted to density profiles \cite{kinast,bartenstein,perali}, collective
excitations \cite{kinast,bartenstein,stringari,combescot,
minguzzi,heiselberg,kim,manini},
condensate fraction \cite{zwierlein,ortiz,salasnich,giorgini}, free
expansion \cite{ohara,bourdel,diana} and vortices \cite{bulgac,ketterle}.

An interesting issue is the inclusion of Bose atoms, like $^{23}$Na or 
$^{87}$Rb isotopes, in the Fermi cloud. Trapped boson-fermion mixtures, with
Fermi atoms in a single hyperfine state, have been investigated by various
authors both theoretically \cite%
{molmer,nygaard,stoof,pethick,viverit1,viverit2,das,liu,wang} 
and experimentally \cite{roati,modugno,ospelkaus}. 
Very recently McNamara {\it et al.} \cite{mcnamara} have reported 
the observation of simultaneous quantum degeneracy 
in a completely spin polarized dilute 
gaseous mixture of $^{3}$He and $^{4}$He in their first excited 
metastable states \cite{nota} 

The purpose of this paper is to carry out a study of the miscibility of a
Fermi-Bose mixture with the Fermi atoms in two equally populated hyperfine
states across a Feshbach resonance of the Fermi-Fermi scattering length.
Since the density of spin up and spin down fermions is bound to be the
same, this three-component system behaves effectively as a two-component
mixture: a superfluid Fermi gas and a  Bose-Einstein condensate.

First we consider the mixture in a box. 
For a positive Bose-Fermi scattering length 
we show that, depending on the total number densities $n_{F}$ and 
$n_{B}$ of the Fermi and Bose  components, 
there are three possible minimum energy 
configurations: a uniform phase with bosons and fermions fully mixed, a 
purely fermionic phase coexisting with a purely bosonic one, and a purely 
fermionic phase coexisting with a mixed phase. We find that the region in 
the $n_{F}$ - $n_{B}$  plane where each one of the three is stable, 
strongly depends on the effective Fermi-Fermi scattering length. 
In addition, we find that when the 
Bose-Fermi scattering length is negative, 
collapse is driven by the dynamical instability of the homogeneous mixture. 

We discuss also the mixture under harmonic
confinement. In this case the conditions for partial demixing, full demixing
and collapse crucially depend on the sign of the Bose-Fermi scattering
length, the ratio between the Bose-Fermi and the Bose-Bose scattering
lengths, and the number of atoms involved. We analyze the experimental
implications of our results by considering mixtures of $^{6}$Li--$^{23}$Na
and $^{40}$K--$^{87}$Rb atoms as examples.

\section{Renormalized scattering length}

It has been recently shown \cite{kim,manini,montecarlo,carlson} that at zero
temperature the energy density of a uniform and dilute Fermi superfluid,
composed by two equally populated spin states, can be written in the BCS-BEC
crossover as 
\begin{equation}
{\cal E} = {\frac{3}{5}} A \, n_F^{5/3} f(y) \; \label{ff-en},
\end{equation}
where $A=\hbar^2(3\pi^2)^{2/3}/(2m_F)$ and $m_F$ is the mass of a fermionic
atom. The universal function $f(y)$ depends on the inverse interaction
parameter $y=(k_F a_F)^{-1}$, where $a_F$ is the Fermi-Fermi scattering
length, $n_F$ is the Fermi number density and $k_F=(3\pi^2 n_F)^{1/3}$ is
the Fermi wave vector. In the experiments on the BCS-BEC crossover the
scattering length $a_F$ is changed by using an external magnetic field
(Feshbach resonance technique) and may be varied from large negative to
large positive values \cite{ohara,greiner,jochim,bourdel}. In these
experiments the effective range of the interaction between fermions is much
smaller than the mean interparticle distance and so the Fermi cloud may be
considered as dilute. In the weakly attractive regime ($y\ll -1$) there is a
BCS Fermi gas of weakly bound Cooper pairs and the universal function $f(y)$
has the asymptotic behavior $f(y)=1 + 10/(9\pi y)+O(1/y^2)$ \cite%
{kim,manini,montecarlo}. In the so-called unitarity limit ($y=0$) one
expects that the energy per particle is proportional to that of a
non-interacting Fermi gas with a coefficient $f(0)=0.42$ \cite{baker}. In
the weak-coupling BEC regime ($y\gg 1$), there is instead a weakly repulsive
Bose gas of molecules of mass $m_M=2m_F$ and density $n_M=n_F/2$. Such
Bose-condensed molecules interact with a positive scattering length $a_M=0.6
a_F$ \cite{montecarlo,petrov} and the universal function $f(y)$ has the
asympotic behavior $f(y)=5 a_M/(18 \pi a_F y) + O(1/y^{5/2})$ \cite%
{kim,manini,montecarlo}. 
Actually the formula $a_M=0.6 a_F$ has been analytically derived
for a weakly-interacting gas of dimers \cite{petrov} but
it seems to work well also in the strongly-interacting limit \cite{montecarlo}. 

In the present work we investigate the effect of Bose atoms mixed with the
two-hyperfine-state gas of fermions by using 
the two different analytical forms Eq. (\ref{f-kz}) and Eq. (\ref{f-ms}) of 
the universal function $f(y)$ proposed by Kim and Zubarev \cite{kim} and  by
Manini and Salasnich \cite{manini} to fit the numerical results 
from Monte Carlo (MC) simulations \cite{montecarlo,carlson} 
and obeying the above asymptotic expressions.

The formula proposed by Kim and Zubarev 
is a [2/2] Pad\`e approximant 
\beq
f(y) = d_0 + {\frac{d_1 y + d_2 }{y^2 + d_3 y + d_4}} \; ,  
\label{f-kz}
\eeq
where the values of the five parameters 
$d_i$ are reported in Ref. \cite{kim}. 
Manini and Salasnich slightly improved the fit 
to the available MC data \cite{montecarlo,carlson} by using a 
different parametrization, i.e. 
\beq
f(y) = \alpha_1 - \alpha_2 \arctan{\left( \alpha_3 y {\frac{\beta_1 + |y| }{%
\beta_2 +|y|}} \right)} \; ,  \label{f-ms}
\eeq
with the values of the parameters $\alpha_i$ and $\beta_j$ reported in
Ref. \cite{manini}. 

\begin{figure}[tbp]
\centerline
{\includegraphics[height=2.5in,clip] {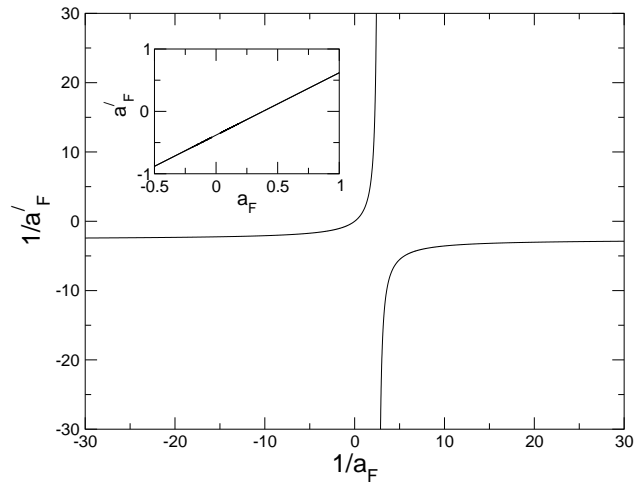}}
\caption{Inverse renormalized scattering length $1/a_F'$
as a function of the inverse bare scattering length $1/a_F$,  
according to Eq. (\ref{ren-af}). 
In the inset there is linear plot of $a_F'$ as a function
of $a_F$. Lengths are in units of $g_{BF}^2/(A g_B)$. }
\end{figure}

At zero temperature a dilute gas of bosons, 
with density $n_{B}$, mass $m_{B}$ 
and interaction strength $g_{B}$, is a Bose-Einstein condensate, with a 
coherence length $\xi _{B}$ given by $\xi _{B}=\hbar /(2m_{B}c_{B})$,
where $c_{B}=(g_{B}n_{B}/m_{B})^{1/2}$ is the bosonic sound velocity. Note
that the strength $g_{B}$ is given by $g_{B}=4\pi \hbar ^{2}a_{B}/m_{B}$,
where $a_{B}$ is the Bose-Bose scattering length. Already back in the 
1960's, in connection with dilute solutions of $^{3}$He in superfluid $^{4}$%
He, it was realized that \cite{pines} the Fermi-Fermi scattering length is
renormalized by the density fluctuations of the bosons and that this
renormalization may lead to a BCS state even when the Fermi-Fermi bare
interaction is repulsive. A thorough analysis 
\cite{stoof,pethick,viverit1,viverit2} of the renormalization 
of the Fermi-Fermi scattering length in Fermi-Bose mixtures 
of ultracold atomic gases has been provided in recent years. 
Viverit \cite{viverit2} has shown that, quite remarkably, 
in the dilute mixture and under the condition 
$k_{F}\xi _{B}\ll 1$, the renormalized scattering length 
of the Fermi-Fermi inter-atomic potential does not 
depend on the densities of the atomic species but only on their interaction
strengths according to 
\beq
{a_{F}^{\prime}}=a_{F}-\frac{m_F}{4\pi \hbar^{2}}
\frac{g_{BF}^2}{g_B} \Theta(n_b)\; ,  
\label{ren-af}
\eeq
where $g_{BF}=4\pi \hbar ^{2}a_{BF}/m_{BF}$ is the Bose-Fermi interaction
strength, $a_{BF}$ its scattering length and $%
m_{BF}=2m_{B}m_{F}/(m_{B}+m_{F})$ the reduced mass. Heaviside's function 
$\Theta(n_b)$ merely states that the renormalization 
occurs only where $n_B \neq 0$. 
\par 
In Fig. 1 we plot the renormalized $1/a_{F}^{\prime }$ as a 
function of $1/a_{F}$ according to Eq. (\ref{ren-af}). 
Fig. 1 shows that the Bose-Fermi interaction
strength $g_{BF}$ drives the system into the non-interacting regime ($%
1/a_{F}^{\prime }=\pm \infty $) when $1/a_{F}=g_{BF}^{2}m_{F}/(g_{B}4\pi
\hbar ^{2})=(8/(9\pi )^{1/3})(Ag_{B}/g_{BF}^{2})$. As previously stressed,
the renormalization formula (\ref{ren-af}) holds if $k_{F}\xi_{B}\ll 1$, 
which may be satisfied only when $g_{B}$ is positive since 
it is equivalent to 
\beq
n_{F}\ll \left( \frac{2 m_B g_B}{m_F A}\right)^{3/2}\; n_{B}^{3/2}\;.
\label{eq5}
\eeq
In the following we will use adimensional densities: 
for fermions $\bar{n}_{F}
=n_{F}\,g_{BF}^{6}/(A^{3}g_{B}^{3})$ and for bosons 
$\bar{n}_{B}=n_{B}\,g_{BF}^{5}/(A^{3}g_{B}^{2})$, 
so that Eq. (\ref{eq5}) becomes 
$\bar{n}_{F}\ll \alpha \,\bar{n}_{B}^{3/2}$, where 
$\alpha=(2 m_B g_B/(m_F |g_{BF}|))^{3/2}$. 
In the case of a mixture of $^{6}$Li
and $^{23}$Na atoms, where $g_{BF}/g_{B}\simeq 0.84$ \cite{ospelkaus}, 
one finds $\alpha\simeq 27$, 
while for a mixture of $^{40}$K and $^{87}$Rb atoms, 
where $g_{BF}/g_{B}\simeq -4.56$ \cite{hadzibabic}, one gets 
$\alpha \simeq 1$. In both cases the range of densities 
where this inequality is valid is sufficiently large to allow experimental 
tests on the predictions of the following sections. 
Note that measuring lengths in units 
of $g_{BF}^2/(A g_B)$, Eq. (4) becomes 
$a_F^{\prime}= a_F - (9\pi)^{1/3}/8 \simeq a_F - 0.38$; 
working near the Feshbach resonance ($a_F'=a_F=\infty$)  
the renormalization of the Fermi-Fermi scattering length is very small. 
\par 
The Eq. (\ref{ren-af}) is valid for both negative and positive $a_F$ 
under the condition (\ref{eq5}) provided that a Fermi-Fermi 
bound state is not formed \cite{viverit2}. 
We expect that the effect of renormalization is negligible close to 
the resonance also on the BEC side of the BCS-BEC crossover, where indeed 
there is the molecular bound state made of two Fermi atoms 
in different hyperfine states. Thus, studying the BCS-BEC crossover, 
in the BEC side ($a_F>0$) we take $a_F'=a_F$ and work 
only close to the Feshbach resonance. 
Of course this issue deserves further investigations 
which are beyond the scope of the present paper. 

\section{Homogeneous Fermi-Bose mixture}

We consider a dilute mixture of $N_B$ bosons and $N_F$ fermions in 
a box of volume $V$. 
As previosly stated, the fermions are equally distributed in two
hyperfine states and can be considered as a single superfluid Fermi
system whose energy density is given by Eq. (\ref{ff-en}). 
Here and in the following we will use a local density 
approximation 
and assume that the interaction parameter $y$ entering the universal
function $f(y)$ depends on the renormalized s-wave 
scattering length between fermions 
$y=(k_F a_F^{\prime})^{-1}$. In our treatment of the BCS-BEC 
crossover we use $a_F'$ given by Eq. (\ref{ren-af}) for $a_F<0$ 
and $a_F'=a_F$ for $a_F>0$. 
The energy density of a uniform, homogeneously mixed phase of 
condensate bosons and superfluid fermions can then be written as 
\begin{equation}
{\cal E} = {\frac{3}{5}} A n_F^{5/3} f(y) + {\frac{1}{2}} g_B \, n_B^2 +
g_{BF} \, n_B \, n_F \; ,  \label{energy-den}
\end{equation}
where $(1/2) g_B n_B^2$ is the energy density 
of a Bose-Einstein condensate while 
$g_{BF}n_B n_F$ is the contribution from the Bose-Fermi interaction. 
$a_B$ and $a_{BF}$ are not significantly modified by the medium 
if the diluteness
conditions $a_B n_B^{1/3}\ll 1$ and $a_{BF}n_F^{1/3} \ll 1$ are satisfied 
\cite{viverit1,viverit2,albus,giorgini3}, therefore we use for them 
their bare values.

The mixed phase is energetically stable if its energy is a minimum
with respect to small variations of the densities $n_F$ and $n_B$, while the
total number of fermions and bosons are held fixed. 
To get the equilibrium densities one must then minimize the function 
\begin{equation}
\tilde{{\cal E}} = {\cal E} - \mu_F \; n_F - \mu_{B} \, n_B \; ,
\end{equation} 
where $\mu_F$ and $\mu_B$ are Lagrange multipliers (imposing that the
numbers of fermions and bosons are fixed) which may be identified with the
Fermi and Bose chemical potentials. Setting the derivatives of $\tilde{{\cal %
E}}$ with respect to $n_F$ and $n_B$ equal to zero, one finds:
\begin{equation}
\mu_F = A \, n_F^{2/3} \left( f(y) - {\frac{1}{5}} f^{\prime}(y) \right) +
g_{BF} \, n_B \; ,  \label{chem-f}
\end{equation}
\begin{equation}
\mu_B = g_B \, n_B + g_{BF} \, n_F \; .  \label{chem-b}
\end{equation}
The solution of Eqs. (\ref{chem-f}) and (\ref{chem-b}), 
gives a minimum if the corresponding Hessian 
of $\tilde{{\cal E}}$ is positive, i.e. if: 
\begin{equation}
{\frac{\partial^2 \tilde{{\cal E}} }{\partial n_F^2}} {\frac{\partial^2 
\tilde{{\cal E}} }{\partial n_B^2}} - \left( {\frac{\partial^2 \tilde{{\cal E%
}} }{\partial n_F \partial n_B}} \right)^2 > 0 \; ,
\end{equation}
implying: 
\begin{equation}
n_F < \left({\frac{2}{3}}\right)^3 \left( {\frac{A \, g_B }{g_{BF}^2}}
\right)^3 \left( f(y) - {\frac{3}{5}} y f^{\prime}(y) + {\frac{1}{10}} y^2
f^{\prime\prime}(y) \right)^3 \; .  \label{ineq}
\end{equation}
(Notice that $n_F$ appears also in the right 
side of the inequality via $y=((3\pi^2 n_F)^{1/3}a_F^{\prime})^{-1}$).
The solution of this inequality gives the region 
in the parameters' space where the homogeneous mixed 
phase is dynamically stable. Dynamical stability, 
corresponding to a local minimum of $\tilde{{\cal E}}$ is a necessary but not
sufficient condition for the energetic stability (also called 
thermodynamical stability at non-zero temperatures) which requires 
instead the global minimum of $\tilde{{\cal E}}$. In fact, we shall show 
that there could exist inhomogeneous, 
two-phase, configurations with energy lower than that
of the uniform, one-phase, configuration considered
in Eq. (\ref{energy-den}). In Fig. 2 we
plot the region of dynamical stability of the homogeneous mixture in the
plane $(1/a_F^{\prime},n_F)$, where $a_F'$ is given by 
Eq. (\ref{ren-af}) for $a_F<0$ and $a_F'=a_F$ for $a_F>0$. 
The solid line corresponds to the parametrization 
of Eq. (\ref{f-kz}) for $f(y)$ in solving the equality in (\ref{ineq}), 
while the dashed line follows from  Eq. (\ref{f-ms}).

Fig. 2 shows that in the BCS regime ($1/a_F^{\prime}\ll -1$) the critical
density $n_F$ is close to $(2/3)^3$. Near the unitarity limit ($%
1/a_F^{\prime}=1/a_F=0$) the critical density 
strongly reduces and the homogeneous
mixture is no longer stable for sufficiently large 
$1/a_F^{\prime}>0$. This is not surprising because 
for a positive $a_F^{\prime}$, where $a_F'=a_F$, 
the Fermi component becomes a gas of Bose-condensed 
molecules of mass $2m_F$, which 
likes to separate from the atoms of mass $m_B$. As shown in Fig. 2, there
exists an upper critical value of $1/a_F^{\prime}$ for the stability of the
homogeneous mixture. 
The existence of such upper critical value can be explained 
by the following simple analytical argument: 
in the deep BEC regime one treats a homogeneous Bose-Bose mixture as
composed of atoms with mass $m_B$ and molecules with mass $m_M=2m_F$ and
interaction strength $g_M=4\pi\hbar^2 a_M/m_M$ with $a_M=0.6 a_F^{\prime}$ 
but $a_F'=a_F$. 
Then, from the stability condition $g_{BF}^2 < g_B g_M/4$, one finds the
critical value at vanishingly small fermion density 
as $1/a_F=1/a_F^{\prime}= \pi/(3\pi^2)^{2/3} (A g_B/g_{BF}^2) = 0.20
(A g_B/g_{BF}^2)$. This is exactly the value where both solid and dashed
lines of Fig. 2 meet the $n_F=0$ axis.
Moreover, the asymptotic behavior 
of the universal function 
$f(y)\sim 1/y + O(1/y^{5/3})$ for large $y$, 
implies that the boundary of the stability region in Fig. 2 
reaches the axis $n_f=0$ with a positive slope. 
As a consequence there must be an interval of  $1/a_F^{\prime}$ values 
where the stability region has both a lower and an upper bound in $n_F$. 
Numerically, one finds that this reentrant behaviour of the stability 
line occurs for values between $0.20 (A g_B/g_{BF}^2)$ 
and $0.28 (A g_B/g_{BF}^2)$ by using Eq. (\ref{f-ms}) 
(see the dashed line in the inset of Fig. 2)) and between 
$0.20 (A g_B/g_{BF}^2)$ and $0.22 (A g_B/g_{BF}^2)$, 
but with much smaller Fermi densities using 
Eq. (\ref{f-kz}) (see the solid line). 

\begin{figure}[tbp]
\centerline
{\includegraphics[height=2.4in,clip] {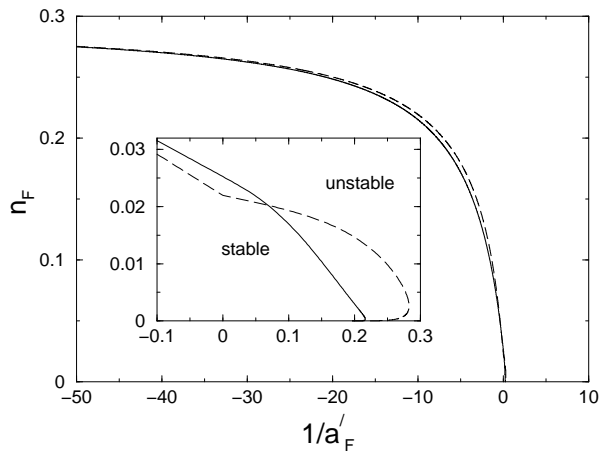}}
\caption{Region of dynamical stability of the homogeneous mixture
in the plane $(1/a_F',n_F)$.
Solid line: obtained with the universal
function $f(y)$ derived by Kim and Zubarev \cite{kim}.
Dashed line: obtained with the universal function $f(y)$ derived by
Manini and Salasnich \cite{manini}. In the inset there is a zoom
of the region with positive Fermi-Fermi scattering length $a_F'$, 
where $a_F'=a_F$. The fermionic density $n_F$ 
is in units of $A^3 g_B^3/g_{BF}^6$ and the scattering length 
$a_F'$ is in units of $g_{BF}^2/(A g_B)$. }
\end{figure}

It is interesting to observe that the inequality (\ref{ineq}) can be written
as 
\begin{equation}
n_F < {\frac{g_B m_F}{g_{BF}^2}} c_F^2 \; ,  \label{ineq2}
\end{equation}
where $c_F=v_F \left(f(y) - {\frac{3}{5}} y f^{\prime}(y) + {\frac{1}{10}}
y^2 f^{\prime\prime}(y)\right)^{1/2}/\sqrt{3}$ is the sound velocity of the
superfluid Fermi component, with $v_F=\sqrt{2 A n_F^{2/3}/m_F}$ the Fermi
velocity. The sound velocity $c_{BF}$ of the Fermi-Bose mixture can be
easily obtained following a procedure similar to the one suggested in Ref. 
\cite{kabanov} for a two-component Bose-Einstein condensate. One finds: 
\begin{equation}
c_{BF}= {\frac{1}{\sqrt{2}}} \sqrt{c_B^2+c_F^2 \pm \sqrt{(c_B^2-c_F^2)^2 + {%
\frac{4 g_{BF}^2 }{m_B m_F}} n_B n_F }} \; ,
\end{equation}
where $c_B=\sqrt{g_B n_B/m_B}$ is the sound velocity of the Bose gas. Thus
the sound velocity has two branches and the homogeneous mixture becomes
dynamically unstable when the lower branch becomes imaginary.

\section{Two-phase Fermi-Bose mixture}

The Bose and Fermi components can form distinct phases, which we label by
the index $i$. If we ignore interpenetration effects, a possible
phase-separated configuration is described by the number $I$ of phases
present, the bosonic densities $n_{B,i}$, the fermionic densities $n_{F,i}$
in each phase, and the fractions $v_i$ of the total volume they occupy.
Since the total number of particles is given, the following relations must
hold: 
\begin{equation}
n_F = \sum_{i=1}^I n_{F,i}v_i \; , \quad \quad n_B = \sum_{i=1}^I n_{B,i}v_i
\; ,
\end{equation}
and $\sum_{i}^I v_i=1$. When $I=1$ we have the case of a homogeneous mixture
discussed in the previous section. The other possiblity for the system we
are considering is $I=2$ \cite{nota1}. In this case the total energy is the
sum of the contributions due to fermions, to bosons and to their mutual
interaction: 
\begin{equation}
{\cal E}= \sum_{i=1}^2 v_i \, \left[ \frac{3}{5}A \, n_{F,i}^{5/3} \, f_i + 
\frac{1}{2} g_B \, n_{B,i}^2 + g_{BF} \, n_{B,i} n_{F,i} \right] \; ,
\label{energia-boh}
\end{equation}
with $f_i=f(y_i)$ and $y_i=((3\pi^2 n_{F,i})^{1/3}a_F^{\prime})^{-1}$, 
where $a_F'$ is given for $a_F<0$ by Eq. (\ref{ren-af}) properly 
renormalizing the scattering length between fermions 
in regions where there are bosons. For $a_F>0$ we set $a_F'=a_F$. 
Note that we have omitted contributions from interface effects 
since they should be negligible if 
the system is in a large box \cite{stchui}. Equilibrium requires the
equality of the pressures  in each phase: 
\begin{equation}
P_i= \frac{2}{5} A \, n_{F,i}^{5/3} \left(f_i - {\frac{1}{2}} y_i
f_i^{\prime}\right) + \frac{1}{2} g_B \, n_{B,i}^2 + g_{BF} \, n_{B,i} \,
n_{F,i} \; ,
\end{equation}
where $f_i^{\prime}=f^{\prime}(y_i)$ , and moreover the
equality of the chemical potentials of each species: 
\begin{equation}
\mu_{F,i} = A\, n_{F,i}^{2/3} \left( f_i -{\frac{1}{5}} y_i
f_i^{\prime}\right) + g_{BF} \, n_{B,i} \; .
\end{equation}
for fermions and:
\begin{equation}
\mu_{B,i} = g_B \, n_{B,i} + g_{BF}\, n_{F,i} \; .
\end{equation}
for bosons.
If the boson density $n_{B,i}$ is non-zero in both phases, then the chemical
potentials $\mu_{B,i}$ must be equal. If the density of one species vanishes in one phase, then its chemical potential in that phase must be higher than in the other. For example, if the boson density is zero  in one
phase, then the boson chemical potential in that phase must be higher than
in the other one, i.e. 
\begin{equation}
\mu_{B,i} > \mu_{B,3-i} \quad \mbox{ if  } n_{B,i} = 0 \; .
\end{equation}
The same is of course true for fermions.

Thus, one has to distinguish four cases of equilibrium, which must be
analyzed one by one:

\noindent (i) {\em Two pure phases}: The bosons and fermions are completely
separated corresponding to $n_{F,1}=0$, $n_{B,2}=0$ and $n_{B,1}\neq 0$, $%
n_{F,2}\neq 0$.

\noindent (ii) {\em A mixed phase and a purely fermionic one}: The boson
density vanishes in one region corresponding to $n_{F,1}\neq 0$, $n_{B,2}=0$
and $n_{B,1}\neq 0$, $n_{F,2}\neq 0$.

\noindent (iii) {\em A mixed phase and a purely bosonic one}: The fermion
density vanishes in one region corresponding to the conditions $n_{F,1} = 0$%
, $n_{B,2}\neq 0$ and $n_{B,1}\neq 0$, $n_{F,2}\neq 0$.

\noindent (iv) {\em Two mixed phases}: All densities $n_{B,1}$, $n_{F,1}$
and $n_{B,2}$, $n_{F,2}$ are different from zero, while $n_{B,1}\neq n_{B,2}$
and $n_{F,1}\neq n_{F,2}$.

\vspace{.1cm}

The analysis of these four cases is similar to that performed for the
Fermi-Bose mixture with a single fermionic hyperfine state by Viverit,
Pethick and Smith \cite{viverit1} and also by Das \cite{das}, who
investigated the one-dimensional mixture. In our problem the presence of the
universal function $f(y)$ increases the complexity of the algebric
manipulations.

We leave out the algebra and discuss the 
results as a function of the inverse scattering length. 
First of all, the analysis shows that the cases (iii) and (iv) are not 
realizable. Thus, a separation into two phases, each with different, 
non-zero, concentrations of bosons and fermions is never in equilibrium, nor
is one with a purely bosonic phase and a mixed one. The cases (i) and (ii)
are instead possible if $g_{BF}>0$, while for $g_{BF}<0$ these
configurations are unstable.

\begin{figure}[tbp]
\centerline
{\includegraphics[height=3.8in,clip] {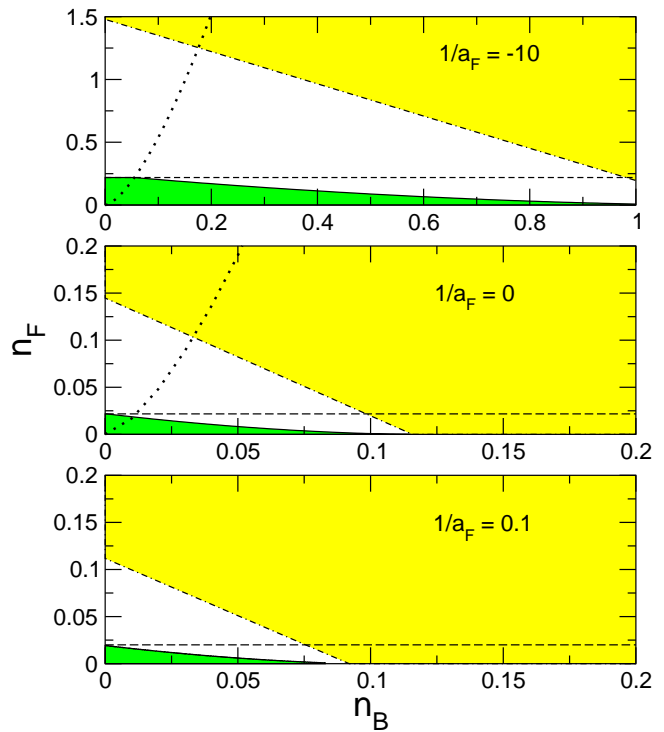}}
\caption{(color online). Phase diagrams in the plane $(n_B,n_F)$
of the Fermi-Bose mixture with $g_{BF}>0$ and three values
of the inverse Fermi-Fermi scattering length $a_F$.
Note the different scales of the top panel. 
Above the dot-dashed line: bosons and 
fermions are completely separated (light grey region).
Between the dot-dashed line and the solid line:
a mixed phase and a purely fermionic one. Below the solid line:
bosons and fermions are fully mixed (dark grey region).
The dotted line gives the curve below which the Eq. (\ref{ren-af})
can be used (mixture of $^6$Li and $^{23}$Na atoms). 
For $a_F>0$ we set $a_F'=a_F$. The bosonic density $n_B$ 
is in units of $A^3 g_B^2/|g_{BF}|^5$. 
The fermionic density $n_F$ is in units of $A^3g_B^3/g_{BF}^6$. 
The Fermi-Fermi scattering length $a_F$
is in units of $g_{BF}^2/(A g_B)$. }
\end{figure}

In Fig. 3 we plot the phase diagrams $(n_B,n_F)$ of the Fermi-Bose mixture
for three values of the bare s-wave scattering length $a_F$ 
and $g_{BF}>0$. In the figure there are the results obtained by using 
Eq. (\ref{f-ms}) to model the universal function $f(y)$. 
Very similar results are 
obtained by using Eq. (\ref{f-kz}). Above the dot-dashed line there are two 
pure phases where bosons and fermions are completely demixed and therefore
$a_F'=a_F$. Between the dot-dashed line and 
the solid line there is a mixed phase with a 
fermion scattering length renormalized by the presence of the bosons 
and a purely fermionic one where $a_F'=a_F$. 
Below the solid line  bosons and 
fermions are homogeneously mixed.
Fig. 3 shows that, as $1/a_F$ is increased from a negative value to 
the Feshbach resonance ($1/a_F=1/a_F^{\prime}=0$), the region 
of pure separation increases, 
while the other two are strongly reduced. 
This effect is stronger above the Feshbach resonance ($1/a_F>0$ and 
$a_F'=a_F$), where pairs of Fermi 
atoms become Bose-condensed molecules. As expected from the dynamical 
stability analysis of the previous section, when $1/a_F$ 
exceedes a critical value the region of 
pure separation occupies the whole phase 
diagram. This critical value is $1/a_F=0.28 (A g_B/g_{BF}^2)$ 
according to Eq. (\ref{f-ms}) and it is instead $1/a_F =0.22 
(A g_B/g_{BF}^2)$ using the Eq. (\ref{f-kz}). 

\begin{figure}[tbp]
\centerline
{\includegraphics[height=1.7in,clip] {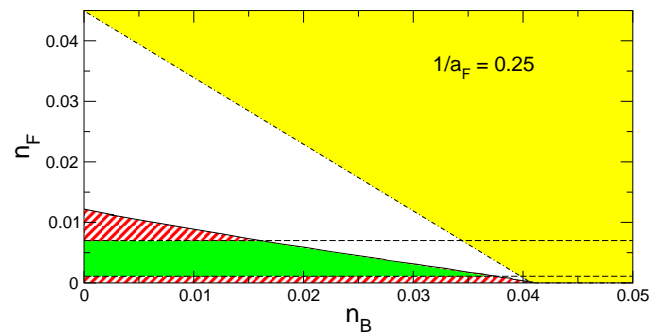}}
\caption{(color online). Phase diagram with $g_{BF}>0$ 
and $1/a_F=0.25$. 
Above the dot-dashed line: bosons and fermions are completely
separated (light grey region). 
Between the dot-dashed line and the solid line:
a mixed phase and a purely fermionic one coexist.
Only in the dark grey region bosons and fermions
are homogeneously mixed. 
In the hatched regions none of the possible
configurations is dynamically and energetically stable. 
Units as in Fig. 3.}
\end{figure}

As discussed in the previous section, a physical consequence of the 
asymptotic behavior of $f(y)$ for $y\to +\infty$ is the existence of an 
interval of $1/a_F$, where the region of dynamical stability of the
mixed phase has not only an upper bound in the fermion density but also a
lower bound. By using Eq. (\ref{f-ms}) this happens for $0.20\,
(Ag_B/g_{BF}^2 ) \le 1/a_F \le 0.28 \, (Ag_B/g_{BF}^2)$. In Fig. 4
we plot the phase diagram obtained with Eq. (\ref{f-ms}) and 
$1/a_F=0.25$. As in Fig. 3, the dark grey region of the diagram is
where the fully mixed configuration is dynamically and energetically stable.
In Fig. 4 there are domains, indicated as hatched regions, where none of the
possible configurations is dynamically and energetically stable. Obviously,
in these domains the total energy is no longer given by Eq. (\ref%
{energia-boh}) and the role of the interface may be relevant. 
There is a similar effect for $g_{BF}<0$. 
In fact, for $g_{BF}<0$ only the single homogeneous
phase, where bosons and fermions are fully mixed, is stable: in Fig. 3 and
in Fig. 4, this region is in between the two dashed lines, and it is
independent on $n_B$. In the other regions of the phase diagram all the
configurations are unstable, and presumibly this corresponds to a collapse
where fermions and bosons are clumped together \cite{viverit1}. 

Our predictions can be verified by using alkali-metal atoms. For the
Fermi-Bose mixture of $^{6}$Li atoms and $^{23}$Na atoms one has $a_{BF}=30
\, a_0$, where $a_0=0.53 \times 10^{-10}$ m is the Bohr radius \cite%
{hadzibabic}. In this case it is easy to find that 
the scaling units of $n_F$ 
and $n_B$, i.e. $(A g_B/g_{BF}^2)^3$ and $A^3 g_B^2/|g_{BF}|^5$, are both
about $10^{8}$ $\mu$m$^{-3}$. For the Fermi-Bose mixture of $^{40}$K atoms
and $^{87}$Rb atoms one has instead $a_{BF}=-284\, a_0$ \cite{ospelkaus} and
the scaling units are about $10^{2}$ $\mu$m$^{-3}$ for fermions and about $%
10^{3}$ $\mu$m$^{-3}$ for bosons. From these densities and Fig. 3 it follows
that in the unitarity limit ($1/a_F=0$) the mixing-demixing
transition can be obtained for instance 
in a mixture of $\sim 10^6$ $^6$Li atoms and $%
\sim 10^6$ $^{23}$Na atoms in a volume of $ \sim 0.5$ $\mu$m$^3$. 
These numbers of atoms are already realized in experiments with ultra-cold 
alkali-metal atoms. 

\section{Inclusion of an external potential}

In actual experiments, the ultra-cold atoms are usually trapped by an
external harmonic potential. For simplicity we consider the same spherical
harmonic potential 
\begin{equation}
U(r) = {\frac{1}{2}} K r^2
\end{equation}
acting on fermions and bosons, where $K=m_F\omega_F^2=m_B\omega_B^2$ is the
elastic constant and $\omega_F$ and $\omega_B$ are the corresponding
harmonic frequencies. The energy functional of the mixture can be written as 
\begin{eqnarray}
E = \int \Big\{  {\frac{\hbar^2 }{2 m_F}} |\nabla \sqrt{n_F(r)}|^2 + {\frac{%
\hbar^2 }{2 m_B}} |\nabla \sqrt{n_B(r)}|^2 \nonumber\\
+ {\cal E}\left(n_F(r),n_B(r)\right) + U(r)\, n_F(r) + U(r)\, n_B(r) \Big\} %
d^3{\bf r}
\end{eqnarray}
where ${\cal E}\left(n_F(r),n_B(r)\right)$ is given by Eq. 
(\ref{energy-den}) while the gradient terms 
give the quantum pressure for bosons and the von
Weisz\"acker contribution to the kinetic energy of fermions with a
nonuniform density. By minimizing the energy with a fixed number $N_B$ of
bosons and $N_F$ of fermions we get 

\[
\mu_F = -{\frac{\hbar^2 }{2 m_F}} {\frac{\nabla^2 \sqrt{n_F(r)}}{\sqrt{n_F(r)%
}}}
\]
\beq
+ A \, n_F(r)^{2/3} \left( f(y(r)) - {\frac{1}{5}} f^{\prime}(y(r))
\right)
+ g_{BF} \, n_B(r) + U(r),  \label{LDA-f}
\end{equation}

\begin{equation}
\mu_B = -{\frac{\hbar^2 }{2 m_B}} {\frac{\nabla^2 \sqrt{n_B(r)} }{\sqrt{%
n_B(r)}}} + g_B \, n_B(r) + g_{BF} \, n_F(r) + U(r) \; ,  \label{LDA-b}
\end{equation}

where $y(r)=((3\pi^2 n_{F}(r))^{1/3}a_F^{\prime})^{-1}$. 
Again $a_F'=a_F$ in the case of full demixing and also for $a_F>0$. 
The presence of a 
harmonic external potential $U(r)$ strongly modifies the conditions for
demixing. Clearly, due to the harmonic confinement partial demixing is
always possible. For instance, with $g_{BF}=0$ both Fermi and Bose clouds
occupy the center of the trap but the effective radii can be quite different
and in this case demixing will appear far from the center of the trap. Such
conditions can be estimated analitically from the previous equations (\ref%
{LDA-f}) and (\ref{LDA-b}) within the Thomas-Fermi (TF) approximation, i.e.
neglecting von Weisz\"acker and quantum-pressure terms. This is a good
approximation when $g_B>0$ and $N_F$ and $N_B$ are large \cite%
{stringari-book,sala-fermi}. For $g_{BF}=0$ the TF approximation gives $%
R_B=\left(15 g_B N_B/(4\pi m_B \omega_B^2)\right)^{1/5}$ for the effective
radius of the Bose cloud. The effective radius $R_F$ of the Fermi cloud
depends instead on $1/a_F^{\prime}$. For $1/a_F^{\prime}\ll -1$ (BCS regime)
one finds $R_F = (48 N_F)^{1/6}\sqrt{\hbar/(m_F\omega_F)}$; for $%
1/a_F^{\prime}=0$ (unitarity limit) one finds $R_F = f(0)^{1/4}(48 N_F)^{1/6}%
\sqrt{\hbar/(m_F\omega_F)}$; for $1/a_F^{\prime}=1/a_F \gg 1$ (BEC regime) 
one finds $R_F=(0.60)^{1/5} \left(15 g_F^{\prime}N_B/(32\pi m_F
\omega_F^2)\right)^{1/5}$, where $g_F^{\prime}=4\pi\hbar a_F^{\prime}/m_F$.

Also for $g_{BF}>0$ partial demixing is possible by varying the number of
atoms. This is the typical situation of the $^6$Li--$^{23}$Na mixture where $%
g_{BF}/g_B= 0.84$. In Fig. 5(a) we plot the density profiles of the Bose and
Fermi clouds obtained in the unitarity limit $1/a_F^{\prime}=0$ with $%
N_B=10^6$ and $N_F=10^5$. In this case the radii of the two clouds are very
similar but, as shown in Fig. 5(b) where $N_B=10^6$ and $N_F=5\times 10^6$,
by increasing the number $N_F$ of fermions one produces partial demixing
with Bose and Fermi atoms in the center of the trap and an external shell of
fermions. The results of Fig. 5 have been obtained by inserting the
Thomas-Fermi density of fermions, $n_F(r)$, into Eq. (\ref{LDA-b}). In this
way one gets the following nonlinear Schr\"odinger equation 
\begin{equation}
\left[ -{\frac{\hbar^2 \nabla^2 }{2 m_B}} + U(r) + g_B n_B(r) + g_{BF}
n_F(r) \right] \psi(r) = \mu_B \psi(r) \; ,
\end{equation}
where $n_B(r)=|\psi(r)|^2$ and $\psi(r)$ is the macroscopic wave function of
the Bose condensate. We have solved this equation by considering its
time-dependent version and using a finite-difference Crank-Nicholson method
with imaginary time \cite{sala-metodi}.

Usually both Fermi and Bose atoms occupy the center of the trap, but under
some conditions one of the two species can be expelled from the center. In
particular, as suggested by Molmer \cite{molmer}, if the bosonic cloud is
practically independent on the fermions, one may get the density profile of
bosons as 
\begin{equation}
n_B(r) \simeq {\frac{1}{g_B}} \left( \mu_B - U(r) \right) \; ,
\end{equation}
by neglecting the quantum-pressure term in Eq. (\ref{LDA-b}). Inserting this
result into Eq. (\ref{LDA-f}) one gets the following TF formula for the
fermionic density profile $n_F(r)$: 
\[
A n_F(r)^{2/3} \left( f(y(r)) - {\frac{1}{5}} f^{\prime}(y(r)) \right)
\simeq 
\]
\begin{equation}
\mu_F - {\frac{g_{BF}}{g_B}} \mu_B - U(r) \left( 1 - {\frac{g_{BF}}{g_B}}
\right) \; ,
\end{equation}
where the terms proportional to $g_{BF}$ are absent in regions with
vanishing $n_B(r)$. This formula shows that if $g_{BF}/g_B<1$ the fermions
will occupy the center of the trap together with bosons. Instead, if $%
g_{BF}/g_B>1$ the fermions are expelled from the center of the trap and form
a shell outside the Bose condensate, i.e there is a core of bosons
surrounded by a shell of fermions. In Fig. 5(c) we show the profiles of the $%
^6$Li--$^{87}$Rb mixture with the artificial value $g_{BF}/g_{B}=5$ and
setting $N_B=10^6$ and $N_F=10^5$. It this case the fermionic cloud is
expelled from the center of the trap. This effect is more clearly shown in
Fig. 5(d) where we set $N_B=10^6$ and $N_F=5\times 10^6$. Note that by using
a much larger value of the ratio $g_{BF}/g_B$ we find full demixing.

\begin{figure}[tbp]
{\includegraphics[height=2.4in,clip]{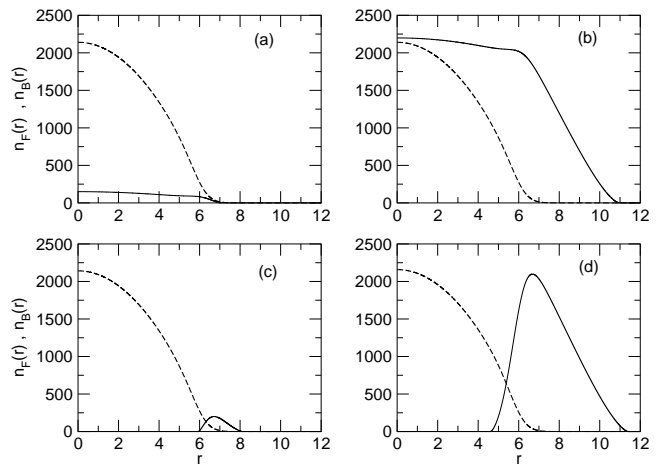}}
\caption{Fermi-Bose mixture of $^6$Li--$^{23}$Na atoms
in a harmonic trap with a resonant Fermi-Fermi
scattering length ($1/a_F=1/a_F'=0$). We set $a_{B}/a_H=0.01$,
where $a_H=\sqrt{\hbar /(m_B\omega_B)}$ is the
characteristic harmonic length of the trap
and $N_B=10^6$. a) $N_F=10^{5}$ and $g_{BF}/g_B=0.84$;
b) $N_F=5\times 10^{6}$ and $g_{BF}/g_B=0.84$;
c) $N_F=10^{5}$ and $g_{BF}/g_B=5$;
d) $N_F=5\times 10^{6}$ and $g_{BF}/g_B=5$.
Solid line: density profile $n_F(r)$ of the cloud of
two-spin $^6$Li atoms; dashed line:
density profile $n_B(r)$ of the cloud of $^{23}$Na atoms.
Lengths are in units of $a_H$ and densities in units of $a_H^{-3}$.}
\end{figure}

For the $^{87}$Rb--$^{40}$K mixture the Bose-Fermi scattering length is
attractive, i.e. $g_{BF}<0$. In this case the Fermi atoms are mixed with the
Bose atoms up to the collapse. As verified by various authors \cite%
{molmer,modugno,miyakawa,capuzzi} with a Fermi-Bose mixture and with
fermions in a unique hyperfine state, the collapse point is well described
by the local density approximation at the point where the dynamical
instability sets up. In our case this point is obtained from Eq. (\ref{ineq}%
) as 
\[
n_F^c(0) = \left({\frac{2}{3}}\right)^3 \left( {\frac{A \, g_B }{g_{BF}^2}}
\right)^3 \left( f(y^c(0)) - {\frac{3}{5}} y^c(0) f^{\prime}(y^c(0)) \right. 
\]
\begin{equation}
\left. + {\frac{1}{10}} y^c(0)^2 f^{\prime\prime}(y^c(0)) \right)^3 \; ,
\end{equation}
where $y^c(0)=((3\pi^2 n_{F}^c(0))^{1/3}a_F^{\prime})^{-1}$ and $n_F^c(0)$
is the critical density at the center of the trap above which there is the
collapse. The plot of $n_F^c(0)$ as a function of $1/a_F^{\prime}$ is
precisely the solid (dashed) curve of Fig. 2 taking $f(y)$ from Eq. (\ref%
{f-kz}) (Eq. (\ref{f-ms})). Using the most recent available data on the $%
^{40}$K--$^{87}$Rb mixture \cite{ospelkaus} we predict that in the deep BCS
regime ($1/a_F^{\prime}\ll -1$) the critical fermionic density is $%
n_F^c(0)=140$ $\mu$m$^{-3}$, while in the unitarity limit ($1/a_F^{\prime}=0$%
) it reduces to $n_F^c(0)=10$ $\mu$m$^{-3}$.

\section{Conclusions}

We have investigated a Fermi-Bose mixture with the Fermi atoms in two
equally populated hyperfine states. In particular, we have analyzed the
mixture as the Fermi-Fermi scattering length is varied across a Feshbach
resonance by using two efficient parametrizations 
of the universal function 
$f(y)$, which characterizes the BCS-BEC crossover. We have found that the
phase diagram of the system strongly depends 
on the Fermi-Fermi scattering length. 
The conditions for demixing and collapse have been studied in a box
and also in a harmonic trap by using a density functional approach. Our
results suggest that in the unitarity limit, where the renormalized 
scattering length of the Fermi-Fermi interaction goes to infinity, demixing
or partial demixing can be observed using a mixture of $^{6}$Li and $^{23}$%
Na atoms. Collapse can be instead obtained using a mixture of $^{40}$K and $%
^{87}$Rb atoms and varying the fermionic density of the system. Many
interesting issues regarding Fermi-Bose mixtures across a Feshbach resonance
remain to be investigated both theoretically and experimentally. Among them
there are finite-temperature effects and population imbalance of fermions in
the two spin states.

L.S. thanks Nicola Manini, Davide Pini and Alberto Parola for many
enlightening discussions.


\begin{thebibliography}{99}
\bibitem{nota} The various superfluid phases
in liquid $^{3}$He-$^{4}$He mixture are
manifestations of quantum degeneracy in dense systems.
\end{thebibliography}
\end{document}